\documentclass[
 reprint,
 amsmath,amssymb,
 aps,
pra,
]{revtex4-2}
\usepackage{silence}
\usepackage{float}
\usepackage{slashed}
\usepackage{xcolor}
\usepackage{array}
\usepackage{amsmath}
\usepackage{subcaption}
\usepackage{multirow}
\usepackage[english]{babel}
\usepackage{booktabs}
\usepackage{lineno, blindtext}
\usepackage[justification=raggedright,singlelinecheck=false]{caption}
\usepackage{tikz}
\usepackage{adjustbox}
\usepackage{rotating}
\usepackage{rotfloat}
\usepackage{diagbox}
\usepackage[utf8]{inputenc}
\usepackage{graphicx}
\usepackage{dcolumn}
\usepackage{bm}

\usepackage{hyperref}
\usepackage{xcolor}
\hypersetup{
  colorlinks   = true, 
  urlcolor     = red, 
  linkcolor    = blue, 
  citecolor   = blue 
}
\usepackage{threeparttable}
\begin{document}
\preprint{APS/123-QED}
\title{Search for a Dark Gauge Boson Within Einstein-Cartan Theory at the ILC Using Multivariate Analysis}

\author{Hossam Taha$^{1,2}$}
\altaffiliation[Hossam.Taha@bue.edu.eg]{}

\author{El-sayed A. El-dahshan$^{2}$}
\altaffiliation{seldahshan@sci.asu.edu.eg}

\author{S. Elgammal$^3$}
\altaffiliation{Sherif.Elgammal@bue.edu.eg}

\affiliation{\\}
\affiliation{$^{1}$\text{Basic Science Department, Faculty of Engineering, The British University in Egypt}}
\affiliation{$^{2}$\text{Physics Department, Faculty of Science, Ain Shams University}}
\affiliation{$^{3}$\text{Centre for Theoretical Physics, The British University in Egypt, P.O. Box 43, El Sherouk City, Cairo 11837, Egypt}}

\begin{abstract}
Multivariate analysis (MVA) is employed to probe the dark matter candidate A$^{\prime}$, a gauge boson of a model rooted within Einstein-Cartan Theory, at the International Linear Collider (ILC). \texttt{WHIZARD} package is used as the event generator to simulate electron-positron interactions at the ILC, at a 500 GeV center-of-mass energy ($\sqrt s$) and a 500 fb$^{-1}$ detector's integrated luminosity ($\mathcal{L}$), to produce the A$^{\prime}$ signal and the expected standard model background. The study focuses on the muonic decay channel of A$^{\prime}$, utilizing several MVA classifiers such as Fisher, Deep Neural Network (DNN), and the Boosted Decision Tree (BDT), aiming to discriminate between the signal of several benchmark points, within theoretical and experimental limits, and the standard model background, and to explore their discovery potential at the ILC. Most benchmark points were proved discoverable at the ILC within $\mathcal{L}= 500$ fb$^{-1}$.

\begin{description}
\item[Keywords]
Beyond the Standard Model, Einstein-Cartan theory, Dark gauge boson, Dark matter, Multivariate analysis, The International Linear Collider (ILC)
\end{description}

\end{abstract}

\maketitle

\section{Introduction}
\label{sec:intro}

The problem is that the Standard Model of Particle Physics (SM) only covers a 5\% of the observable universe. Yes, it succeeds in describing three out of the four fundamental forces. However, it fails in explaining other phenomena, such as those attributed to Dark Matter (DM). \cite{sm1,sm2,DE}.

DM continues to be one of the most fascinating mysteries in Physics, proposed to solve the 85\% unidentified portion of the mass of the universe (about 27\% of the observable universe). It is a type of matter that does no interactions with any existing SM particle. However, it is heavily supported by several astrophysical measurements \cite{dm1,dm2,dm3}. In order to explore it, we would have to look outside the SM's scope, as if it were a lower-energy special case of more comprehensive generalized higher-energy theories, that may potentially succeed in unifying all four fundamental forces \cite{bsm}.

Additional observations in cosmology, astrophysics, and particle physics have inspired the presence of more $U(1)$ gauge bosons that have minimal coupling to particles in the SM. Implying that such bosons might exist in the dark sector of the universe and interact with SM particles via kinetic mixing only \cite{dgp1,dgp2}, hence usually called dark gauge bosons.

Einstein-Cartan theory (ECT) represents a modified framework of general relativity in which the scalar torsion field (ST) that incorporates the spacetime's torsion property allows for extra degrees of freedom \cite{tor1,tor2,tor3,Einstein-Cartan}. 
The ECT framework enables the coupling of gravity, encoded in ST, to all fermions in the SM through the axial-vector mode. This offers a method to investigate DM via studying the dark gauge boson (A$^{\prime}$) that interacts via kinetic mixing with particles of the SM.

DM searches typically involve producing visible SM particles, such as jets or $W,Z$ bosons \cite{R35, atlasMonoZ,CMSmonoz}, photons \cite{photon, photonATLAS} or the Higgs boson \cite{R36, monoHiggsAtlas1, monoHiggsAtlas2}, along with a missing transverse energy ($E_T^{miss}$) accounting for DM candidates. This results in a final state of $SM + E_T^{miss}$.

Previous searches for $Z'$, a massive gauge boson predicted by Supersymmetry and the Grand Unified Theory (GUT) \cite{Extra-Gauge-bosons, gaugeboson1, LR-symmetry11, Super-symmetry12}, were conducted by CMS and ATLAS with no evidence found to support its existence in RUN II of the Large Hadron Collider (LHC), excluding the existence of Z$^{\prime}$ for masses between 0.6 and 5.1 TeV at a 95\% Confidence Level. \cite{zprime, zprimeATLAS}.

The International Linear Collider (ILC) represents a planned linear electron-positron ($ e^{+} e^{-}$) collider, intended for Higgs boson studies. However, it can also be employed to investigate new, beyond the SM, physics \cite{bsmilc}. The $e^{+} e^{-}$ collisions offered by the ILC promise high precision and efficient energy utilization, as they involve elementary particles that can fully exploit the available center-of-mass energy. Additionally, this approach provides a cleaner experimental environment, free from the complex quark interactions typically present in proton-proton collisions. Furthermore, the ILC employs controllable beam polarization and initial particle energy techniques.

We argue that the ILC, at a 500 GeV center-of-mass energy ($\sqrt s$), and a 500 fb$^{-1}$ detector's integrated luminosity ($\mathcal{L}$) \cite{ilc1,ilc2}, would be suitable to explore possible neutral dark gauge boson candidates of low masses.
As they are beyond the capabilities of the LHC in its TeV energy scale, and due to the considered masses of the ST in this study being significantly larger in comparison to the $\sqrt s$ of $\sim$200 GeV of the Large Electron Positron Collider, they would not have been observed there.

In this paper, the ECT model is utilized to probe the A$^\prime$ by simulating the circumstances of the $ e^{+} e^{-}$ interactions at the ILC, at $\sqrt s$ = 500 GeV and $\mathcal{L} =  500$  fb$^{-1}$, via a Monte-Carlo (MC) event generator to discriminate between several signal benchmark points (BMPs) and the SM background, and to study the sensitivity of the ILC to each considered BMP. Performing this study at the ILC provides valuable insights to the high-energy physics community, contributing to the crucial decision-making process regarding the next generation of colliders.

Conventional cut-based analysis often encounters limitations when discriminating subtle signal signatures from irreducible SM backgrounds. To address this, we incorporate multivariate analysis (MVA) techniques, tailored to our signal topology's distinct kinematic and event-shape characteristics to maximize sensitivity to the studied signal BMPs.

The paper starts with the description of the simplified ECT model, which predicts A$^{\prime}$ production, along with its free parameters, in Section \ref{section:model}. Then we present the simulation techniques and methods utilized in generating events for the SM background and the ECT model in Section \ref{section:simulation}. Next, we provide the plan of the MVA, the specific classifiers, and observables used, in Section \ref{section:Analysis}. Followed by presenting the results of this study in Section \ref{section:results}, and then the summary at the end in Section \ref{section:conclusion}.

\section{THE SIMPLIFIED MODEL BASED ON EINSTEIN-CARTAN THEORY}
\label{section:model}
The ECT model presented in \cite{Einstein-Cartan} suggests that ECT could serve as a channel to investigate A$^{\prime}$, where ST can couple to the fermions of the SM.
Within the ECT model, there are two scenarios: the bremsstrahlung scenario, which allows only for one of the DM fermions to radiate A$^\prime$, and the cascade scenario that allows radiation of A$^{\prime}$ from the two DM fermions. This study focuses on the first scenario. 

\begin{figure}
        \centering
        \includegraphics[width=.8\linewidth]{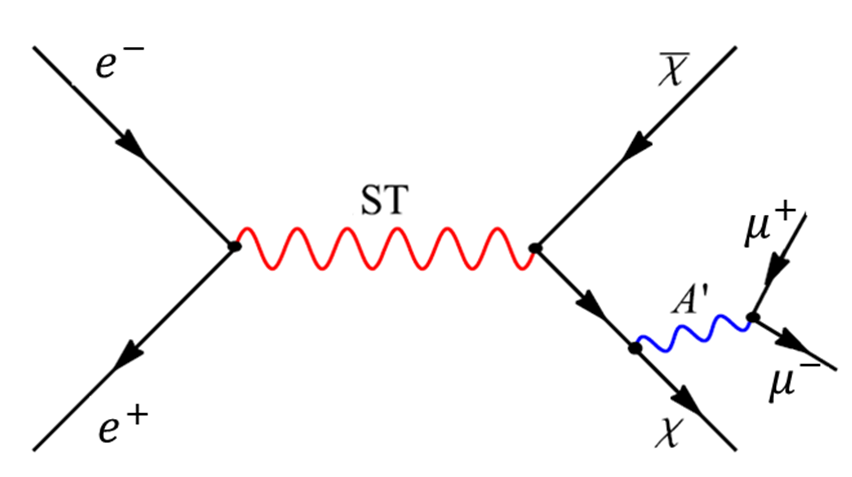}
        \caption{
The topology of the ECT signal - mediated by the torsion field (ST), electron and positron interact with each other to produce two DM fermions, A$^{\prime}$ is then radiated via bremsstrahlung by one of the DM fermions.}
  \label{figure:fig1}
    \end{figure}

The production of A$^\prime$ (or $A^\prime_\mu$) can be achieved at the ILC through the pair annihilation process of $e^+e^-$, mediated by the heavy ST (or $S_\mu$) which accounts for the torsion's axial–vector tensor component ($T^{\lambda}_{\mu\nu}$ ). 
ST then transforms into two DM fermions ($\chi\bar{\chi}$) massive enough to radiate A$^{\prime}$ via bremsstrahlung that decays into di-muon, as illustrated in Figure \ref{figure:fig1}.

The terms responsible for the interaction of the torsion field with the Dirac fermion $\psi$ in the effective Lagrangian are given by \cite{Einstein-Cartan}:

\begin{equation}
\bar{\psi}i\gamma^{\mu}(\partial_\mu+ig_{\eta}\gamma^5S_\mu+...)\psi, 
\label{eqn1}
\end{equation}

where $\texttt{g}_{\eta}$ is the coupling of ST to Dirac fermions, while the coupling of the ST to the DM fermions, and the coupling between the DM fermions and $A^\prime$, is described by the term \cite{Einstein-Cartan}: 

\begin{equation}
\bar{\chi}(i\gamma^{\mu}D_\mu-M_\chi)\chi,
\label{eqn2}
\end{equation}

where $D_\mu=\partial_\mu+ig_{\eta}\gamma^5S_\mu+ig_D A^\prime_\mu$, the coupling of A$^\prime$ to the DM fermion is represented by $\texttt{g}_{D}$, while $M_\chi$ denotes the DM fermion mass. 

We've chosen the di-muon decay channel of the neutral A$^{\prime}$, since it can decay into two SM fermions. The largest possible branching ratio (BR) for $A^\prime\rightarrow\mu^+\mu^-$ is achievable upon satisfying this condition of \cite{Einstein-Cartan}:

\begin{equation}
M_{A^\prime}<2M_\chi.
\label{eqn3}
\end{equation}

The free parameters of the ECT model contain the coupling constants $\texttt{g}_{\eta}$ and $\texttt{g}_{D}$, and the masses of the torsion field ($\text{M}_{ST}$), the DM fermion ($\text{M}_\chi$), the dark gauge boson ($\text{M}_{A^\prime}$). 

The ECT model has primarily investigated the production of dark matter through proton-proton collisions in \cite{Einstein-Cartan}. To adapt this model for the studied $e^+e^-$ collider, we adjusted some parameters. Specifically, we set the coupling $\texttt{g}_{\eta}$ to zero for quarks in order to avoid potential constraints on the torsion field from previous searches at the LHC. For leptons, we set $\texttt{g}_{\eta}$ to 0.125 as taken from \cite{Einstein-Cartan}, abandoning universal coupling in the process (however, non-universal couplings are still well motivated). Meanwhile, we retained the coupling $\texttt{g}_{D}$ at a value of 1.2, as specified in \cite{Einstein-Cartan}.

The CMS collaboration has searched for low mass dilepton resonances ($m<200$ GeV) as outlined in \cite{zprimeCMSlowmass}, successfully excluding dilepton resonance masses in the range of 11.5 - 45.0 GeV. In a previous study, we investigated the ECT model at the ILC that involved analyzing a fixed dark gauge boson (A$^{\prime}$) mass of 10 GeV, which helped establish constraints on other free parameters within the model \cite{dmilc}. Based on the above, we focus on A$^{\prime}$ masses in the range between 50 and 90 GeV. 

In order to guarantee that A$^{\prime}$ is produced with a high boost while satisfying the mass criterion of equation (\ref{eqn3}), we set the DM fermion mass $M_{\chi} =$ 90 GeV. Additionally, we considered ST masses of M$_{ST}$ = 1500 GeV and 2000 GeV.

The expected signature of the process provides a clean channel to SM backgrounds as it contains a muon pair of opposite charges produced via the decay of A$^{\prime}$ accompanied by a large $E_T^{miss}$ taken by the non-decaying DM fermions $\chi\bar{\chi}$, leading the considered events to have a final state of $\mu^+\mu^-+E_T^{miss}$.

\section{Simulated background and signal samples}
\label{section:simulation}

The processes composing the SM background with produced muon pairs in the final state inside the signal domain are the Drell-Yan (DY) process, the di-bosons production ($W^+W^-$ and $ZZ$), and the pair production of top quarks.
Event generation was conducted via the multi-purpose matrix element event generator \texttt{WHIZARD}~v2.8.5 \cite{whiz} for the SM background and ECT signal. To implement the ECT model into \texttt{WHIZARD}, we have requested the necessary files from the author of \cite{Einstein-Cartan}. The modeling of showering and hadronization was done using \texttt{PYTHIA v6} \cite{pyth}, via \texttt{WHIZARD}'s built-in interface with it.
\texttt{DELPHES}~\cite{delph} was employed to simulate the ILC detector's response, for the read-out system response (digitization), and to simulate the reconstruction processes.
All events were generated via $e^+e^-$ collisions at $\sqrt s$ = 500 GeV, and $\mathcal{L}$ = 500 fb$^{-1}$, with polarization degrees of $P_{e^{+}}=0.3$ and $P_{e^{-}}=-0.8$ for the positron and the electron beams respectively, corresponding to the ILC conditions mentioned in \cite{ilc1,ilc2}.

\begin{table}[h]
\centering
\renewcommand{\arraystretch}{1.5}
\begin{tabular} {|c|l|c|}
\hline
\hspace{0.2cm} \textbf{Process} \hspace{0.2cm} & \hspace{0.2cm} \textbf{Final state} \hspace{0.2cm}  & \hspace{0.3cm} \textbf{$\sigma \times \mathrm{BR}$ (fb)} \hspace{0.3cm} \\
\hline
\hline
Drell-Yan & $\mu^+\mu^-$ & 1850 $\pm$ 70  \\
WW  & $\mu^+\mu^- + 2\nu$ & 232.5 $\pm$ 0.6 \\
$t\bar{t}$ & $\mu^+\mu^- + 2b + 2\nu$ & 10.36 $\pm$ 0.02 \\
ZZ & $\mu^+\mu^- + 2\nu$ & 3.72 $\pm$ 0.01 \\
ZZ  & $4\mu$ & 0.470 $\pm$ 0.002 \\
\hline
\end{tabular}
\caption{The processes of the SM background that match the signature of the ECT signal, along with the cross-sections of each process multiplied by the branching ratio at the ILC with $\sqrt{s}$ = 500 GeV and polarization degrees of $P_{e^{-}} = -0.8$, $P_{e^{+}} = 0.3$.}
\label{BKGs}
\end{table}

Table \ref{BKGs} shows the final state of each MC background sample, along with its cross-section ($\times$ BR) computed at a leading order (LO).
Regarding the chosen signal BMPs, the measurements of cross-section ($\times$ BR) were calculated (also at LO) along with the theoretical uncertainty of 1-$\sigma$, calculated on a statistical basis, and displayed in Table \ref{BMPs}.
Systematic and statistical uncertainties were considered for both background samples and signal BMPs.
The contributions of the processes (signal BMPs and SM background) were determined using MC simulations, and were normalized to the ILC's 500 fb$^{-1}$ integrated luminosity and to their cross-sections.
\begin{table}[h]
\centering
\renewcommand{\arraystretch}{1.5}
\begin{tabular} {|c|c|c|c|}
\hline
        \hspace{0.15cm} \textbf{BMPs} \hspace{0.15cm} & \hspace{0.1cm} $\text{M}_{ST}$ (GeV) \hspace{0.1cm} &  \hspace{0.1cm} $\text{M}_{A^\prime}$ (GeV) \hspace{0.1cm} & \hspace{0.1cm} \textbf{$\sigma \times \mathrm{BR}$ (fb)} \hspace{0.1cm} \\
\hline
\hline
       1  & 1500 & 50 & 1.649 \\
       2  & 1500 & 60 & 1.290 \\
       3  & 1500 & 70 & 1.025 \\
       4  & 1500 & 80 & 0.816 \\
       5  & 1500 & 90 & 0.651 \\
       \hline
       \hline
       6  & 2000 & 50 & 0.473 \\
       7  & 2000 & 60 & 0.371 \\
       8  & 2000 & 70 & 0.293 \\
       9  & 2000 & 80 & 0.233 \\
       10 & 2000 & 90 & 0.186 \\

\hline

\end{tabular}

\caption{The ECT signal benchmark points used in this study, along with their production cross-section ($\times$ BR) in fb at LO for multiple values of the A$^{\prime}$ masses ($M_{A'}$) in GeV 
and $ST$ masses ($M_{ST}$) in GeV; and at coupling constants of $\texttt{g}_{\eta} = 0.125,~\texttt{g}_{D} = 1.2$.
at the ILC with $\sqrt{s}$ = 500 GeV and polarization degrees of $P_{e^{-}} = -0.8$, $P_{e^{+}} = 0.3$. The cross-section's uncertainty is about 0.2\%.}
\label{BMPs}
\end{table}

A pre-selection is performed to select events characterized by two oppositely charged muons accompanied by an $E_T^{miss}$, representing the DM fermion candidates ($\chi\bar{\chi}$). The requirements of the selection, presented in Table \ref{cuts}, are applied as restrictions on some kinematic variables. 
\begin{table}[h]
\centering
\renewcommand{\arraystretch}{1.5} 
\begin{tabular} {|c|c|c|}
\hline
\textbf{Step} & \textbf{Variable} & \textbf{Requirement} \\
\hline
\hline

\multirow{3}{*}{Pre-selection}

& $p^\mu_T \ (GeV) $  & $> 10$ \\ 

& $|\eta^\mu| \ (rad) $ & $< 2.4$ \\ 

& IsolationVar & $< 0.1$ \\ 

\hline

\end{tabular}
\caption{The pre-selection criteria used in this study.}
\label{cuts}
\end{table}

$p^\mu_T$ is the momentum of the produced muons in the transverse direction, $|\eta^\mu|$ is the di-muon's pseudo-rapidity, while the third variable, IsolationVar, describes an isolation criterion within \texttt{DELPHES} utilized to exclude muons formed within jets. It ensures that the total of $p_T$ of all relevant tracks around a muon candidate within a cone (with $\Delta R = 0.5$), excluding this muon, is lower than $0.1 p^\mu_T$. The number of events accepted by the selection of each SM background/BMP is presented in Table \ref{preselec}.

\begin{figure}[h]
	\centering
	\includegraphics[width=0.5\textwidth]{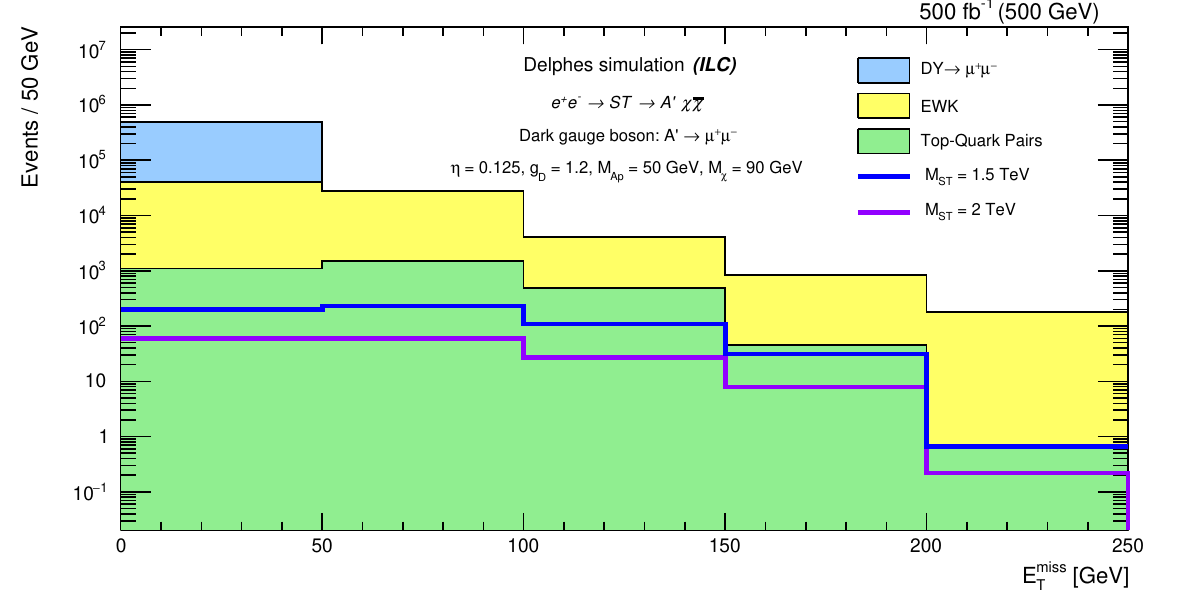}
	\caption{The missing transverse energy distributions for the SM background, alongside the signal benchmarks point (BMPs) 1 and 6, is presented, with $M_{A'}$ set at 50 GeV with coupling constants $\texttt{g}_{D} = 1.2$ and $\texttt{g}_{\eta} = 0.125$. This simulated data reflects events surviving the selection criteria of Table \ref{cuts}.}
\label{METs}
\end{figure}

Figure \ref{METs} illustrates the $E_T^{miss}$ distributions for events accepted by the selection criteria presented in Table \ref{cuts}. The cyan histogram represents the Drell-Yan process ($Z \rightarrow \mu^{+}\mu^{-}$), while the yellow histogram corresponds to the backgrounds from vector boson pairs (ZZ and WW). Additionally, the light-green histogram stands for the $t\bar{t}$ process, all presented in a stacked format. Overlaying these distributions are the analyzed BMPs 1 and 6 at $M_{ST}$ = 1500 and 2000 GeV, respectively, and a $M_{A^\prime}$ of 50 GeV. BMP1 is represented by the non-filled blue histogram, while BMP6 is represented by the non-filled purple histogram.
\begin{table}[h]
\centering
\renewcommand{\arraystretch}{1.5}
\begin{tabular} {|c|c|c|}
\hline
\hspace{0.2cm} \textbf{Type} \hspace{0.2cm} & \hspace{0.2cm} \textbf{Process} \hspace{0.2cm} & \hspace{0.2cm} \textbf{Accepted events} \hspace{0.2cm} \\
\hline
\hline
\multirow{6}{*}{\textbf{SM Background}}
& Drell-Yan  & 449190 $\pm$ 44924  \\
& WW  &  67974 $\pm$ 6802    \\
& $t\bar{t}$ & 3144 $\pm$ 319  \\
& ZZ($2\mu 2\nu$) & 230 $\pm$ 28  \\
& ZZ($4\mu$) & 1259 $\pm$ 131 \\ 
\cline{2-3}
& Total & 521797 $\pm$ 52185 \\
\hline
\hline
\multirow{10}{*}{\textbf{Signal BMPs}}
&1 & 536 $\pm$ 58 \\
&2 & 467 $\pm$ 49 \\
&3 & 369 $\pm$ 42 \\
&4 & 305 $\pm$ 35 \\
&5 & 247 $\pm$ 29 \\
\cline{2-3}
&6 & 154 $\pm$ 20 \\
&7 & 128 $\pm$ 17 \\
&8 & 106 $\pm$ 15 \\
&9 & 87 $\pm$ 13 \\
&10 & 70 $\pm$ 11 \\
\hline
\end{tabular}
\caption{Number of events accepted by pre-selection criteria for each background of the SM and BMP of the ECT signal at $\sqrt{s}=500\ GeV$ and $\mathcal{L}=500\ fb^{-1}$.}
\label{preselec}
\end{table}

\section{the Multivariate Analysis}
\label{section:Analysis}

Machine learning–based multivariate classification techniques have become essential tools in modern high-energy physics analyses as the ongoing search for extremely small signals within large datasets requires extracting, from the data, the largest possible amount of information  \cite{TMVA}.

As illustrated in Figure \ref{METs}, the considered ECT signal BMPs are completely dominated by the SM background. Therefore, we used a MVA, employing \texttt{Fisher}, Deep Neural Network (DNN), and the Boosted Decision Tree (BDT) classifying algorithms available via the toolkit for multivariate analysis package \texttt{(TMVA)} \cite{TMVA} within \texttt{ROOT} framework \cite{ROOT}, to separate between the signal and the SM background. MVA methods outperform traditional cut-based analysis by enabling a more flexible partitioning of the phase space. Decision trees, in particular, divide the phase space into numerous multidimensional hypercubes, each categorized as either background-like or signal-like. This approach allows the creation of non-linear boundaries in hyperspace that enhance the discrimination of the signal and the background. The cut-based analysis, on the other hand, can only select a single hypercube of phase space as the signal region, losing a lot of discrimination power.
The specific parameters of each classifier used in the MVA are tabulated in Table \ref{paras}. \\
\indent BMP6 was selected as the reference point for the signal in all plots moving forward.

\renewcommand{\arraystretch}{1.3}
\begin{table*}[]
    \centering
    \begin{tabular}{|c|c|c|c|}
        \hline
        \textbf{Classifier} & \textbf{Parameter} & \textbf{Value} & \textbf{Description} \\ 
        \hline \hline

        \multirow{5}{*}{\textbf{Fisher}} 
        & VarTransform           & None     & Variable transformation method             \\ \cline{2-4}
        & CreateMVAPdfs          & Enabled  & Create PDFs for MVA output distributions    \\ \cline{2-4}
        & PDFInterpolMVAPdf      & Spline2  & Interpolation method for MVA PDFs           \\ \cline{2-4}
        & NbinsMVAPdf            & 50       & Number of bins for MVA PDFs                 \\ \cline{2-4}
        & NsmoothMVAPdf          & 10       & Number of smoothing iterations for PDFs     \\ \hline \hline

        \multirow{8}{*}{\textbf{DNN}}
        & Layout           & TANH|128, TANH|128 & Network layers and activations               \\ \cline{2-4}
        & LearningRate     & 1e-2     & Learning step size                          \\ \cline{2-4}
        & Momentum         & 0.9      & Gradient momentum                           \\ \cline{2-4}
        & ConvergenceSteps & 20       & Validation steps                            \\ \cline{2-4}
        & BatchSize        & 100      & Samples per training batch                  \\ \cline{2-4}
        & TestRepetitions  & 1        & Test cycle repetition                       \\ \cline{2-4}
        & DropConfig       & 0.0 + 0.5 + 0.5 + 0.5 & Dropout rates per layer \\ \cline{2-4}

        & WeightDecay      & 1e-4     & Weight decay regularization
        \\ \hline \hline

        \multirow{9}{*}{\textbf{BDT}}
        & NTrees           & 1000     & Number of boosting trees                    \\ \cline{2-4}
        & MinNodeSize      & 2.5\%    & Minimum leaf size                           \\ \cline{2-4}
        & BoostType        & AdaBoost & Boosting type                               \\ \cline{2-4}
        & AdaBoostBeta     & 0.5      & Beta value for AdaBoost                     \\ \cline{2-4}
        & UseBaggedBoost   & True     & Use bagging during training                 \\ \cline{2-4}
        & BaggedSampleFraction & 0.5  & Bagged sample fraction                      \\ \cline{2-4}
        & MaxDepth         & 3        & Tree depth                                  \\ \cline{2-4}
        & SeparationType   & GiniIndex& Node splitting metric                       \\ \cline{2-4}
        & Ncuts            & 27       & Number of cuts                              \\ \hline

    \end{tabular}
    \caption{List of hyperparameters for the Fisher, DNN, and BDT methods, along with their descriptions and corresponding values used in the MVA.}
    \label{paras}
\end{table*}

We used the following nine observables as input variables to the three classifying algorithms (Fisher, DNN, BDT):

\begin{enumerate}
    \item ${E}_T^{\text{miss}}$, the missing energy in the transverse direction.
    \item $M_{\mu^+\mu^-}$, The di-muon's invariant mass. 
    \item $\Delta\phi_{\mu^+\mu^-,\vec{E}{_T^{miss}}}$, The azimuthal angle difference between the di-muon system and the missing energy in the transverse direction.
    \item $\Delta R_{\mu^+\mu^-}$, the angular separation of the two muons.
    \item $\eta^{\mu_1}$, the psuedo-rapidity of the first muon.
    \item $\eta^{\mu_2}$, the psuedo-rapidity of the second muon.
    \item $p^{\mu_1}_T$, the momentum of the first muon in the transverse direction.
    \item $p^{\mu_2}_T$, the momentum of the second muon in the transverse direction.
    \item $cos(\theta_{3D})$, the cosine of the 3D angle between the di-muon system vector and the missing transverse energy vector.

\end{enumerate}

\begin{figure*}[]
	\centering
    \begin{subfigure}{0.42\textwidth}
		\includegraphics[width=\textwidth]{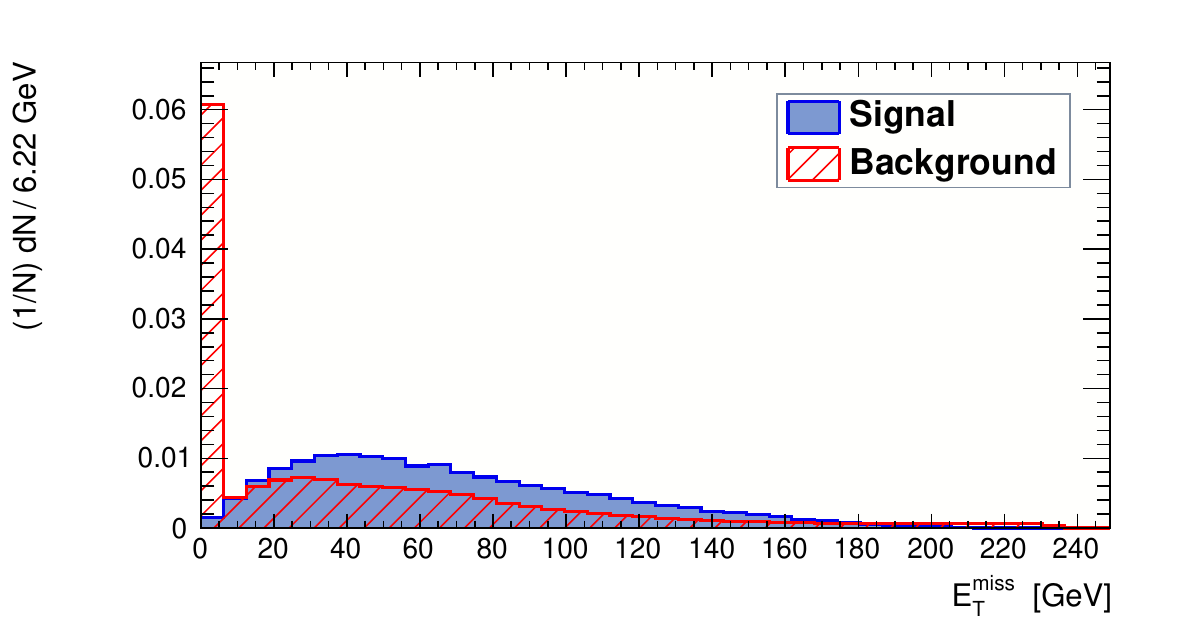}
		\caption{$E^{miss}_T$}
        \label{etmiss}  
	\end{subfigure}
        \begin{subfigure}{0.42\textwidth}
		\includegraphics[width=\textwidth]{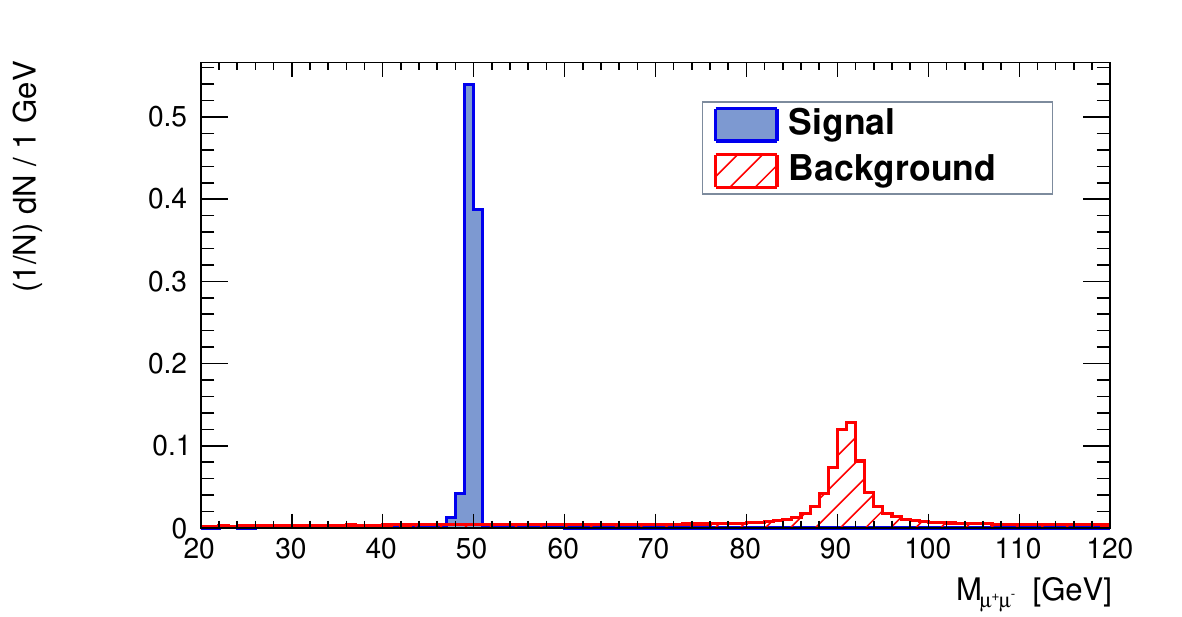}
		\caption{$M_{\mu^+\mu^-}$}
        \label{invmass}  
	\end{subfigure}
    \begin{subfigure}{0.42\textwidth}
		\includegraphics[width=\textwidth]{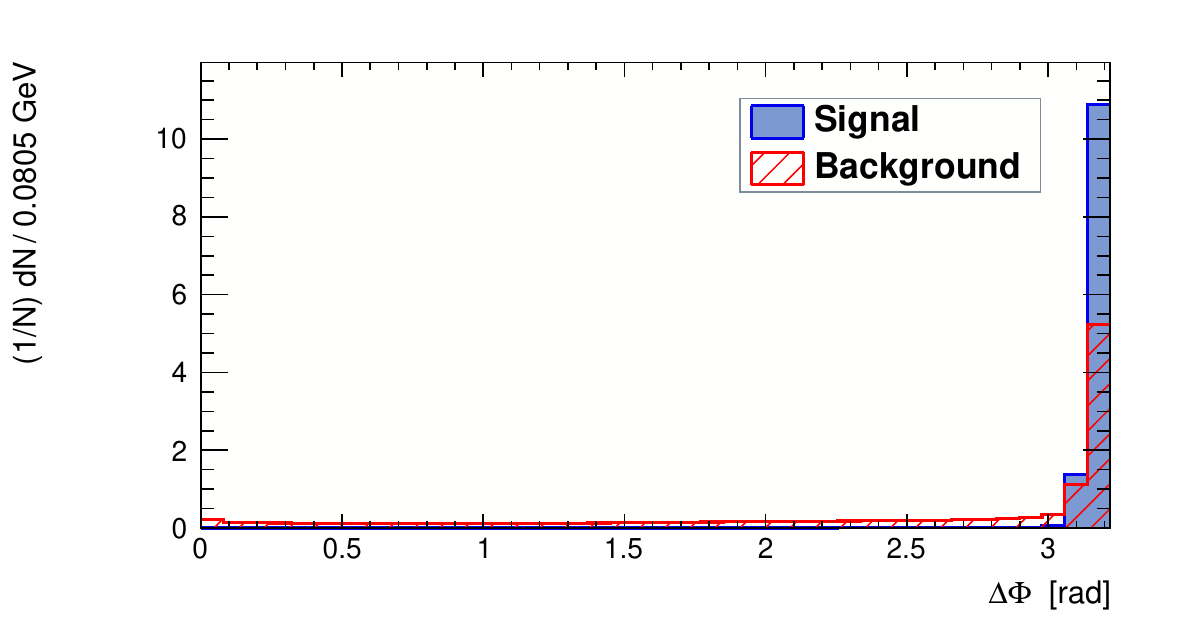}
		\caption{$\Delta\phi$}
        \label{dphi}  
	\end{subfigure}
    \begin{subfigure}{0.42\textwidth}
		\includegraphics[width=\textwidth]{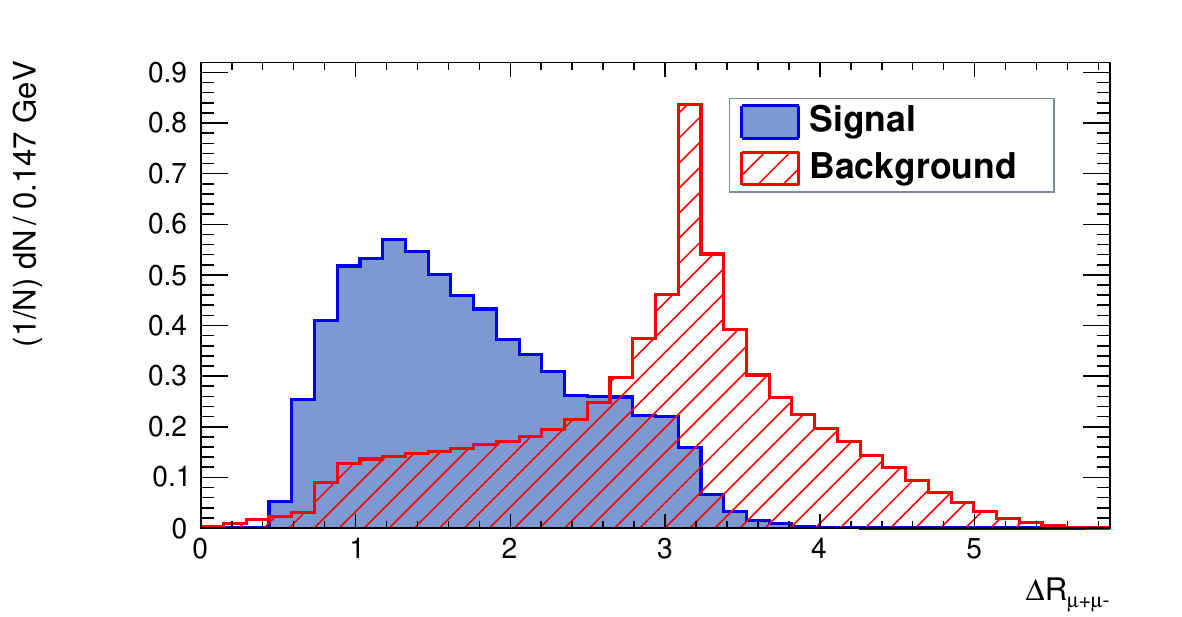}
		\caption{$\Delta R$}
        \label{dR}  
	\end{subfigure}
    \begin{subfigure}{0.42\textwidth}
		\includegraphics[width=\textwidth]{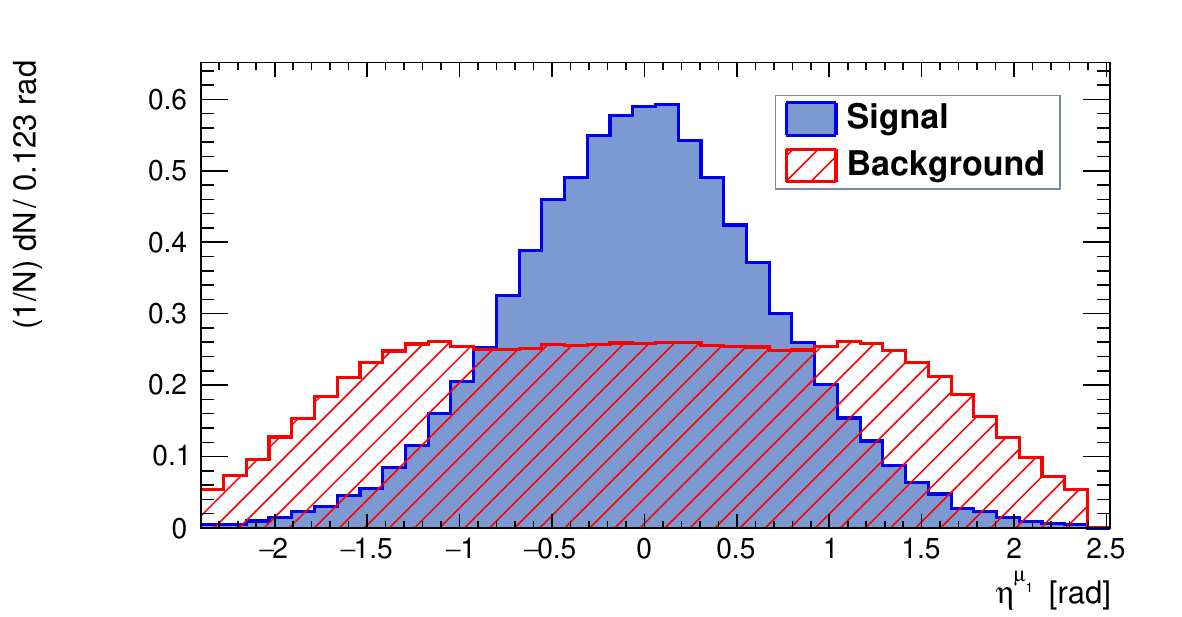}
		\caption{$\eta^{\mu_1}$}
        \label{eta1}  
	\end{subfigure}
    \begin{subfigure}{0.42\textwidth}
		\includegraphics[width=\textwidth]{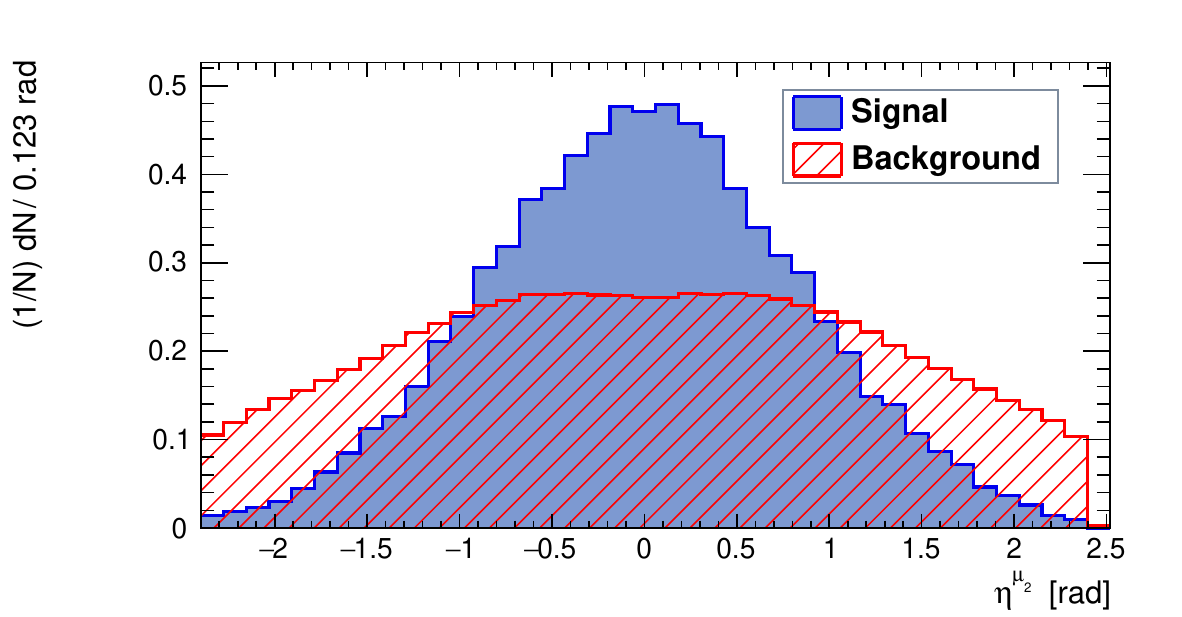}
		\caption{$\eta^{\mu_2}$}
        \label{eta2}  
	\end{subfigure}
    \begin{subfigure}{0.42\textwidth}
		\includegraphics[width=\textwidth]{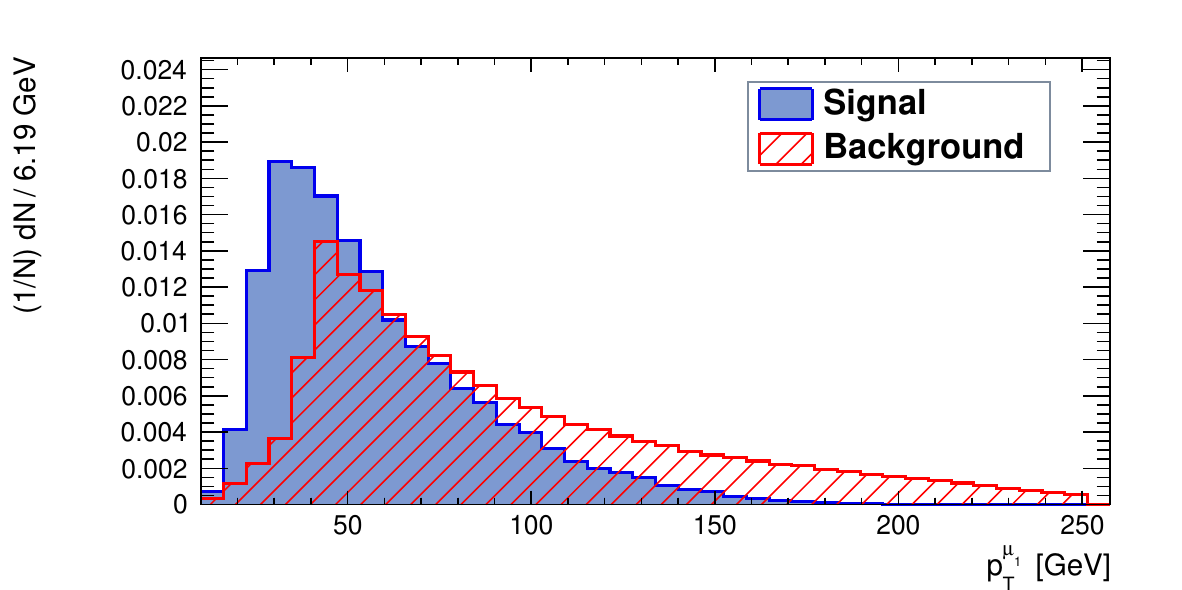}
		\caption{$p^{\mu_1}_T$}
        \label{pt1}  
	\end{subfigure}
	\begin{subfigure}{0.42\textwidth}
		\includegraphics[width=\textwidth]{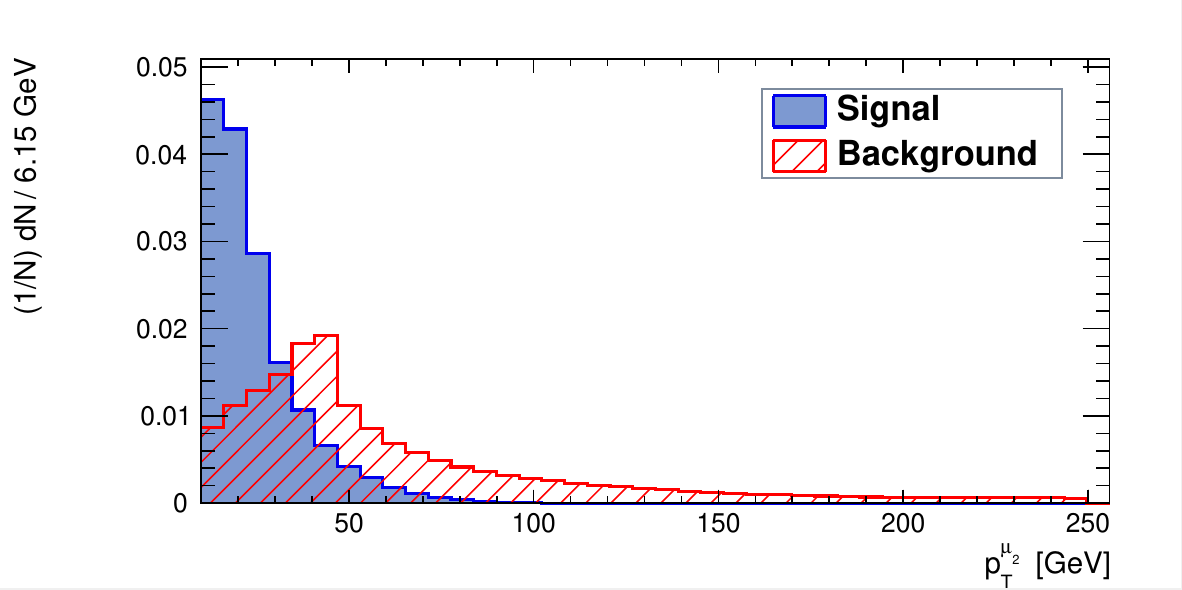}
		\caption{$p^{\mu_2}_T$}
        \label{pt2}
	\end{subfigure}
    \begin{subfigure}{0.42\textwidth}
		\includegraphics[width=\textwidth]{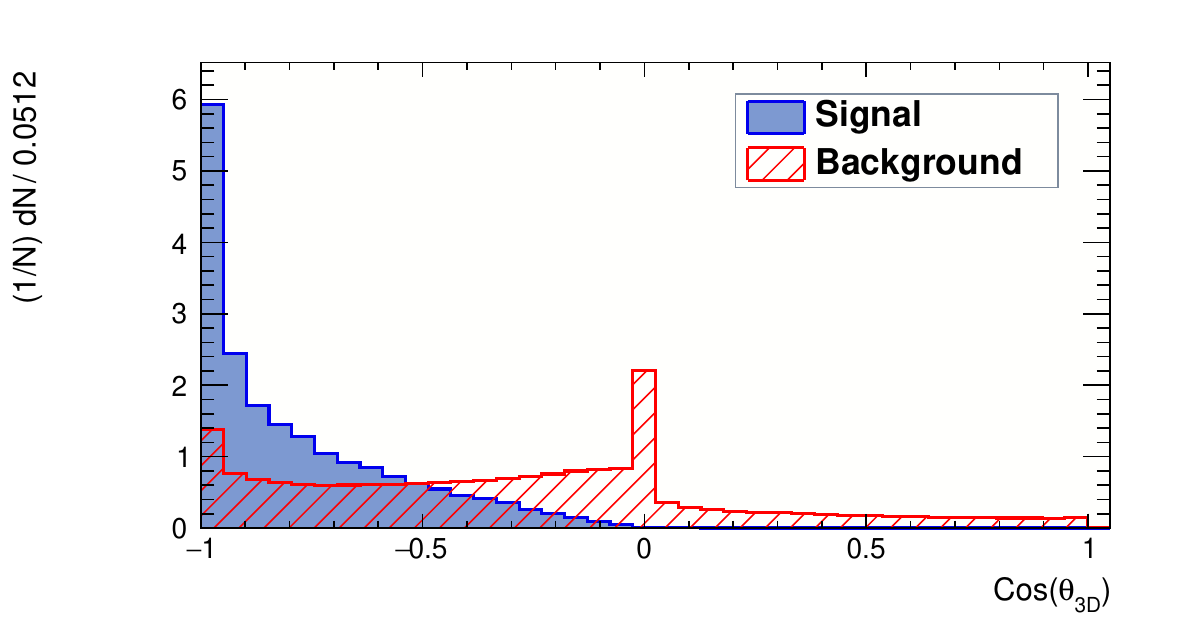}
		\caption{$cos(\theta_{3D})$}
        \label{3D}  
	\end{subfigure}
	\caption{The distributions of the nine input variables employed in the MVA testing processes for the SM background and the signal BMP6 ($\text{M}_{A^\prime}=50$ GeV, $\text{M}_{ST}=2000$ GeV, and $\text{M}_{\chi}=90$ GeV) at $\sqrt s=500$ GeV.}
\label{invars}
\end{figure*}

Figure \ref{invars} shows the normalized distributions of the input variables for the signal BMP6 versus all of the SM background, where the Y-axis represents the bin size.


\begin{figure}[!h]
     
     \centering
     \includegraphics[width=1\linewidth]{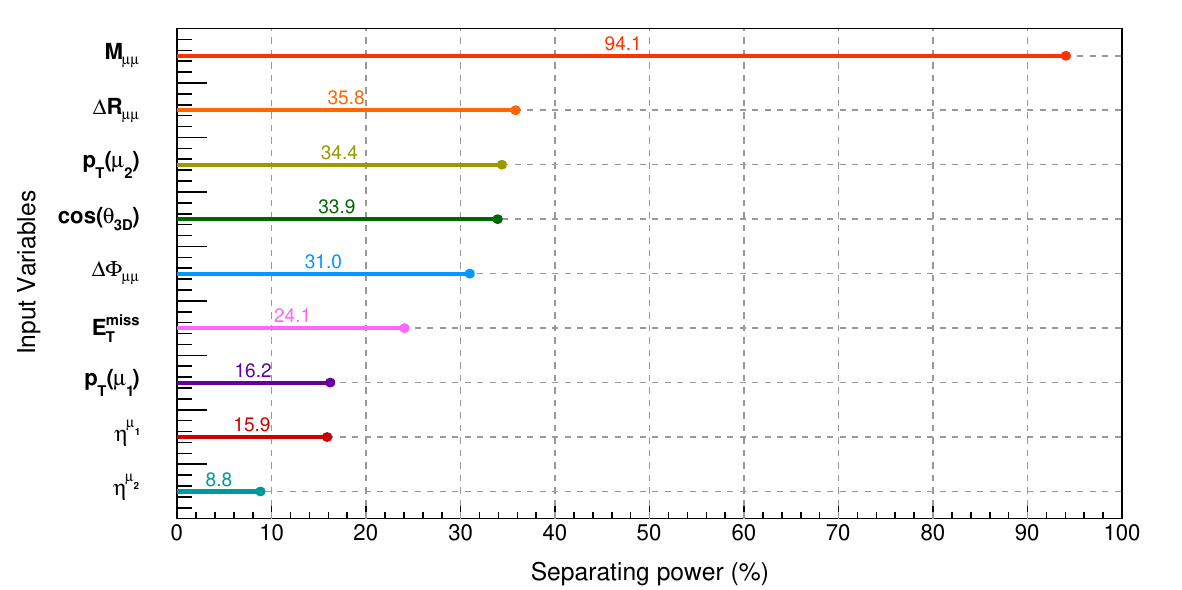}
     \caption{The separating power of each of the input variables employed in the MVA (method unspecific) for BMP6 versus all background at $\sqrt s=500\ GeV$.}
     \label{sepower}
 \end{figure}

\indent The normalized method-unspecific separating power of an observable $\alpha$ is given by \cite{TMVA}:
\begin{equation}
    \mathcal{S}_{(\alpha)}=\frac{1}{2}\int \frac{\left(\hat{f}_{S}(\alpha)-\hat{f}_{B}(\alpha)\right)^{2}}{\hat{f}_{S}(\alpha)+\hat{f}_{B}(\alpha)} d \alpha
    \label{eqn5}
\end{equation}
where $\hat{f}_{S}(\alpha)$ and $\hat{f}_{B}(\alpha)$ are respectively the probability distribution functions of the signal and background for a given observable $\alpha$, the limits of integration correspond to the allowed range of $\alpha$, while $\mathcal{S}_{(\alpha)}$ quantifies the separating power for an observable $\alpha$, and the $\frac{1}{2}$ is used to normalize the values of $\mathcal{S}_{(\alpha)}$ from 0 to 1.
For identical signal and background shapes,
$\hat{f}_{S}(\alpha)=\hat{f}_{B}(\alpha)$ and $\mathcal{S}_{(\alpha)}$=0, while $\mathcal{S}_{(\alpha)}$ = 1 corresponds to a perfect discriminating power. Figure \ref{sepower} illustrates the separating power of each observable used as an input variable in the MVA. 

\begin{figure*}[t]
	\centering
    \begin{subfigure}{0.45\textwidth}
		\includegraphics[width=\textwidth]{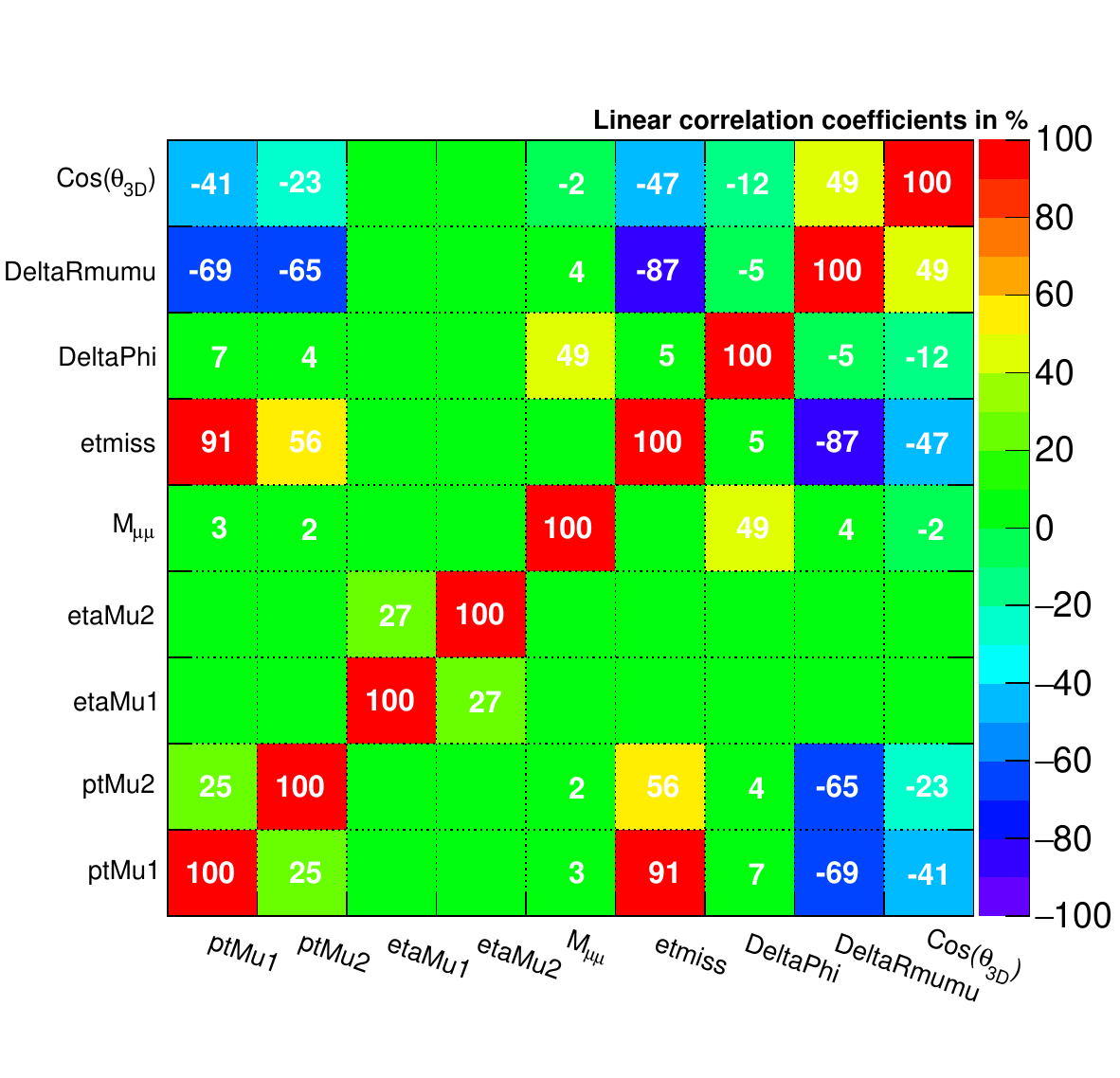}
		\caption{The correlation matrix of the signal BMP6.}
        \label{limit1}  
	\end{subfigure}
	\begin{subfigure}{0.45\textwidth}
		\includegraphics[width=\textwidth]{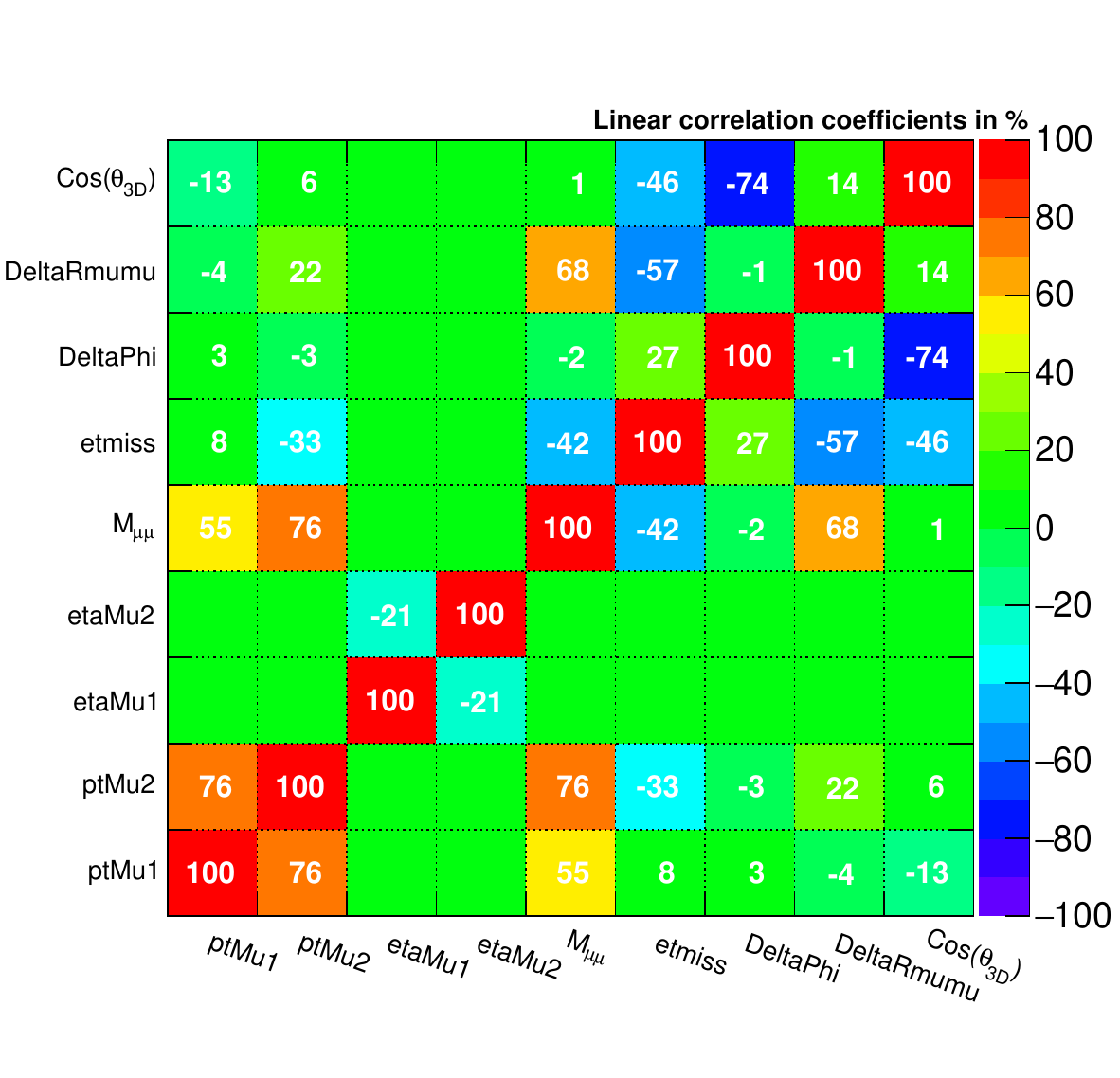}
		\caption{The correlation matrix of the SM background.}
        \label{limit2}
	\end{subfigure}
	\caption{Correlation matrices calculated via the \texttt{TMVA} for signal BMP6 and total SM background at $\sqrt{s}=500$ GeV. Positive and negative values represent correlation and anti-correlation, respectively.}
\label{corrmat}
\end{figure*}

The linear correlation coefficient $\rho$ of two variables $\gamma$ and $\lambda$ is calculated through the Pearson correlation coefficient formula \cite{TMVA}: 
\begin{equation}
\rho_{(\gamma,\lambda)} = \frac{\text{cov}(\gamma,\lambda)}{\sigma_{(\gamma)}\sigma_{(\lambda)}}
\label{eqn4}
\end{equation}
where $\text{cov}(\gamma,\lambda)$ is the covariance of the variables $\gamma$ and $\lambda$, while $\sigma_{(\gamma)}$ and $\sigma_{(\lambda)}$ represent the standard deviations of $\gamma$ and $\lambda$, respectively. 
Linear correlation is a key factor in determining whether a variable contributes independent information. 
Most variables used in the analysis are minimally correlated, as illustrated in Figure \ref{corrmat}.


Positive values of $\rho$ represent correlation, while the negative values represent anti-correlation. 
Certain variables exhibit low correlation in the background while showing moderately high correlation in the signal (and vice versa). This is due to the different kinematics of each process (background and signal). Though minimally correlated variables should be used, they are kept due to their good separating power, which also has a significant contribution within the MVA.
This is usually decided whenever a variable has a distinct correlation for the background and the signal, because after the classifier applies the cut, variables may exhibit reduced correlation in different sectors of phase space \cite{BJT}. 
For example, \texttt{Fisher} event selection is carried out in a transformed space where the input variables have zero linear correlation. The method works by identifying differences in the average values of the background and signal distributions.
The Linear Discriminant Analysis (LDA) finds an axis in the original correlated hyperspace along which background and signal events are separated as much as possible, ensuring that the output classes are maximally separated when projected onto this axis. At the same time, events belonging to the same category (background or signal) are grouped closely together along this axis \cite{TMVA}. 

After using the \texttt{TMVA} to calculate the correlation of the variables employed, the k-fold cross-validation (CV) approach was used to train the classifiers with k = 5 for each BMP and its corresponding SM background. This method divides the dataset into five equal parts (folds). 
The model trains on four folds ($k-1$), and uses the fifth for validation. The procedure runs for k iterations, using each fold as the validation set exactly once \cite{CV1,CV2}.
The CV approach offers a strong evaluation of model performance and helps mitigate over-fitting, especially in HEP analyses, where data imbalance or statistical fluctuations are common.

\begin{figure}[h]
        \centering
        \includegraphics[width=1\linewidth]{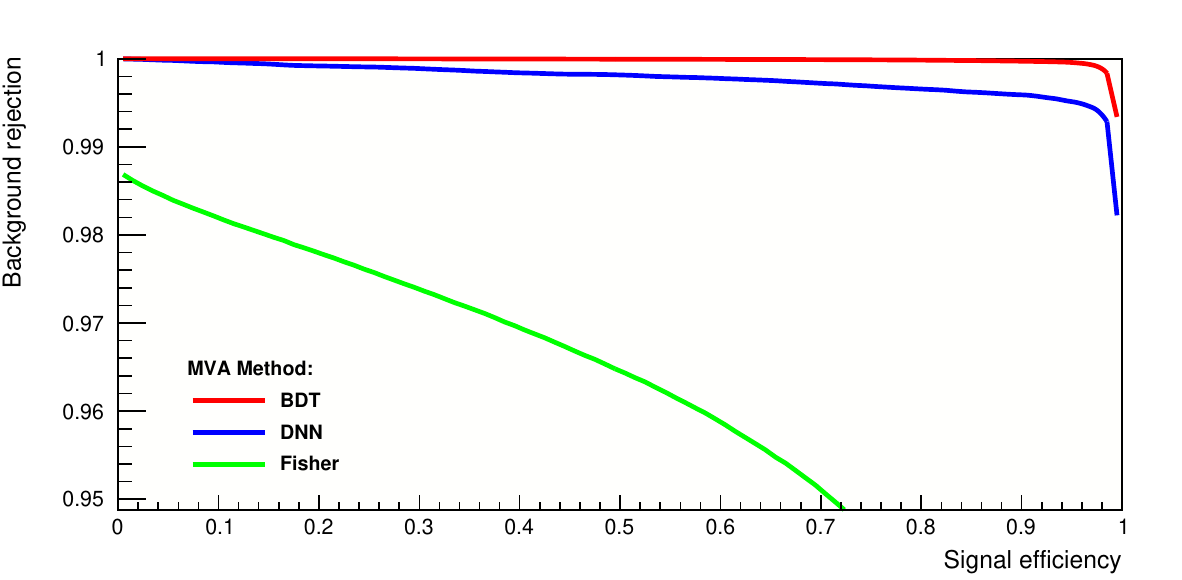}
        \caption{The ROC curves of the three classifiers for signal BMP6 versus all background at $\sqrt s=500\ GeV$, illustrating high background rejection by the BDT method, followed by the DNN and Fisher methods.}
  \label{ROC}
    \end{figure}

Figure \ref{ROC} shows the receiver operating characteristic (ROC) curves for the three classifiers, plotting the background rejection ($1- $ false positive rate) against the signal efficiency (true positive rate) at various threshold settings.
The area under the curve, having values in between 0 and 1, represents a metric for the classifier's performance, as high values of it would represent a better discriminative power. Based on Figure \ref{ROC}, the best classifier is the BDT, followed by the DNN, and then Fisher. Thus, the study proceeds with the BDT classifier.
\begin{figure}[!bh]
	\centering
    \begin{subfigure}{0.42\textwidth}
		\includegraphics[width=\textwidth]{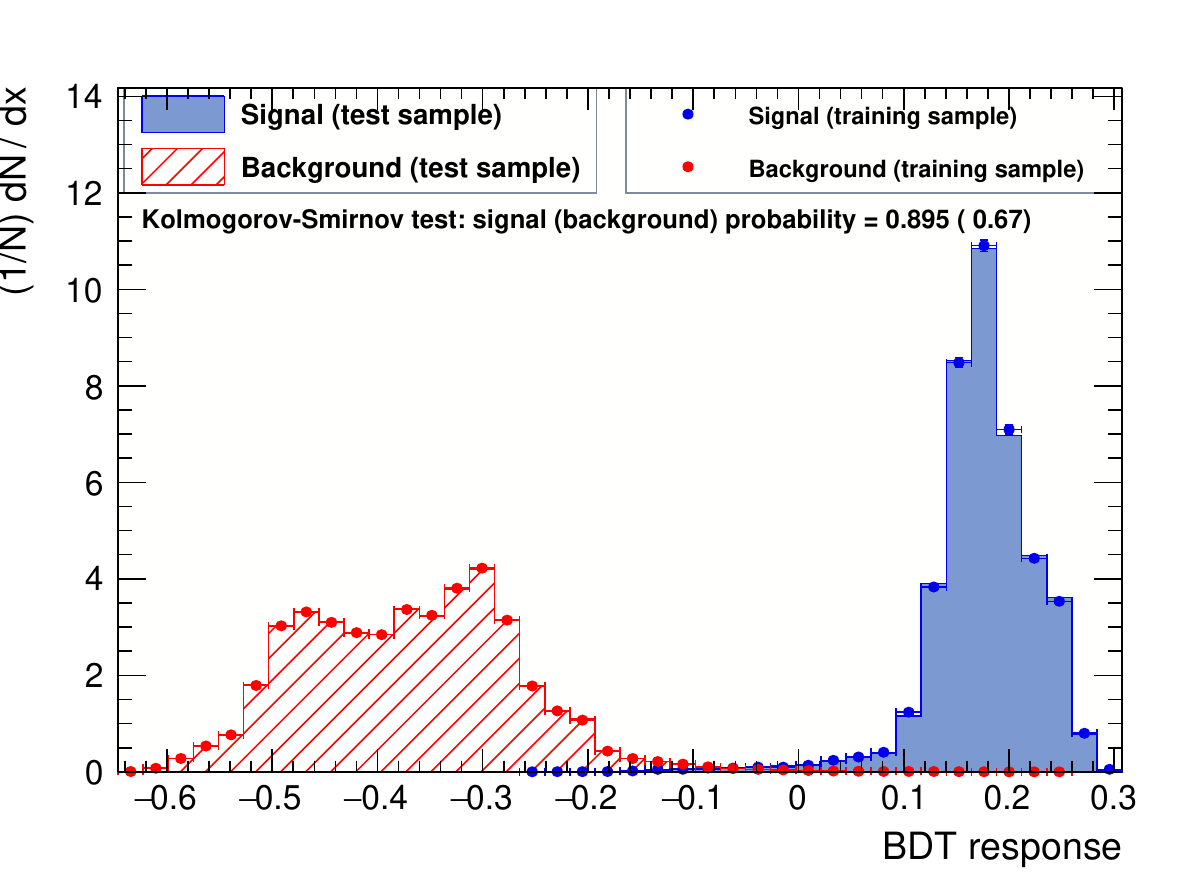}
		\caption{BDT}
        \label{BDT}  
	\end{subfigure}
    \begin{subfigure}{0.42\textwidth}
		\includegraphics[width=\textwidth]{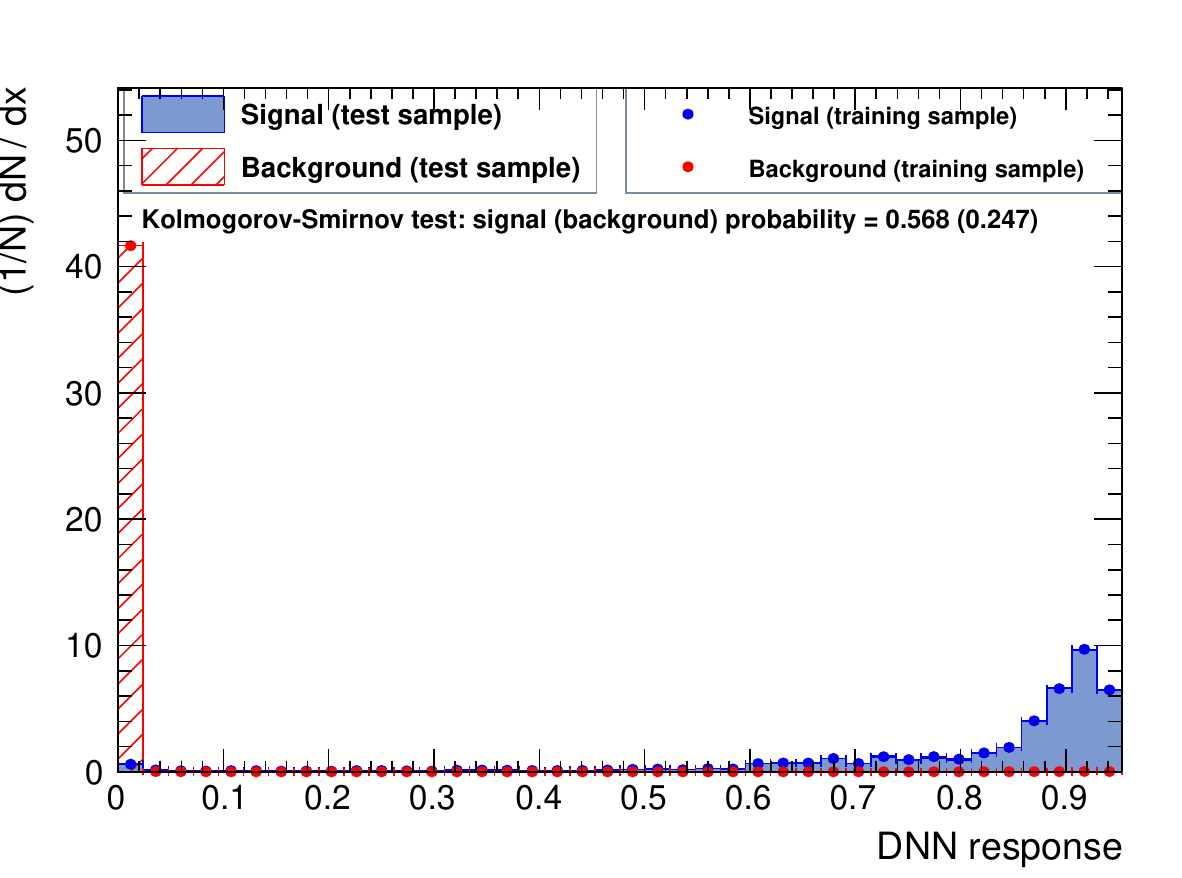}
		\caption{DNN}
        \label{DNN}  
	\end{subfigure}
	\begin{subfigure}{0.42\textwidth}
		\includegraphics[width=\textwidth]{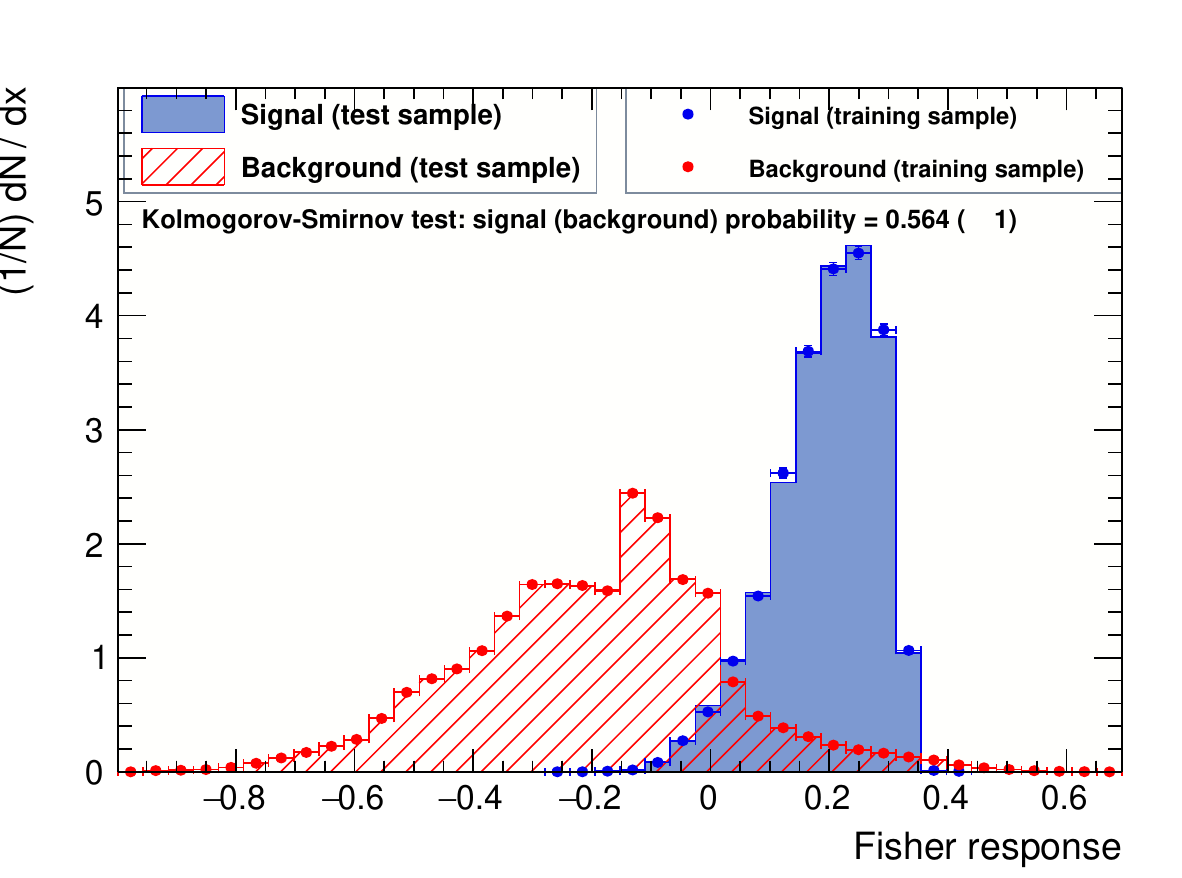}
		\caption{Fisher}
        \label{fisher}
	\end{subfigure}
	\caption{\texttt{TMVA} overtraining check for the three classifiers BDT in \ref{BDT}, 
    DNN in \ref{DNN} and Fisher in \ref{fisher} for SM background and signal BMP6, with
    $M_{A'}$ set at 50 GeV, $M_{ST}$ = 2000 GeV, $M_{\chi}$ = 90 GeV, with coupling constants of $\texttt{g}_{D} = 1.2$ and $\texttt{g}_{\eta} = 0.125$.}
\label{classresp}
\end{figure}

To verify that this result is not due to overtraining, we utilized the \texttt{TMVA} overtraining check illustrated in Figure \ref{classresp} for the three classifiers: BDT (shown in \ref{BDT}), DNN (shown in \ref{DNN}), and Fisher (shown in \ref{fisher}). 
The \texttt{TMVA} response distributions present BMP6 (as the signal) and the SM background, utilizing training and testing samples. The strong alignment of the training and testing histograms for both the signal and the SM background indicates no overtraining, highlighting the effective performance of all three classifiers.  

Quantitatively, the Kolmogorov–Smirnov test (KST) is used to assess potential overtraining \cite{KST}. The KST compares the cumulative distributions (for each classifier’s output) of the training and testing samples, providing a statistical measure of their agreement. A high p-value ($>0.05$) from the KST indicates that the two distributions are statistically similar, confirming that the classifiers generalize well to unseen data.
\section{Results}
\label{section:results}
\begin{figure}[h]
     \centering
     \includegraphics[width=1\linewidth]{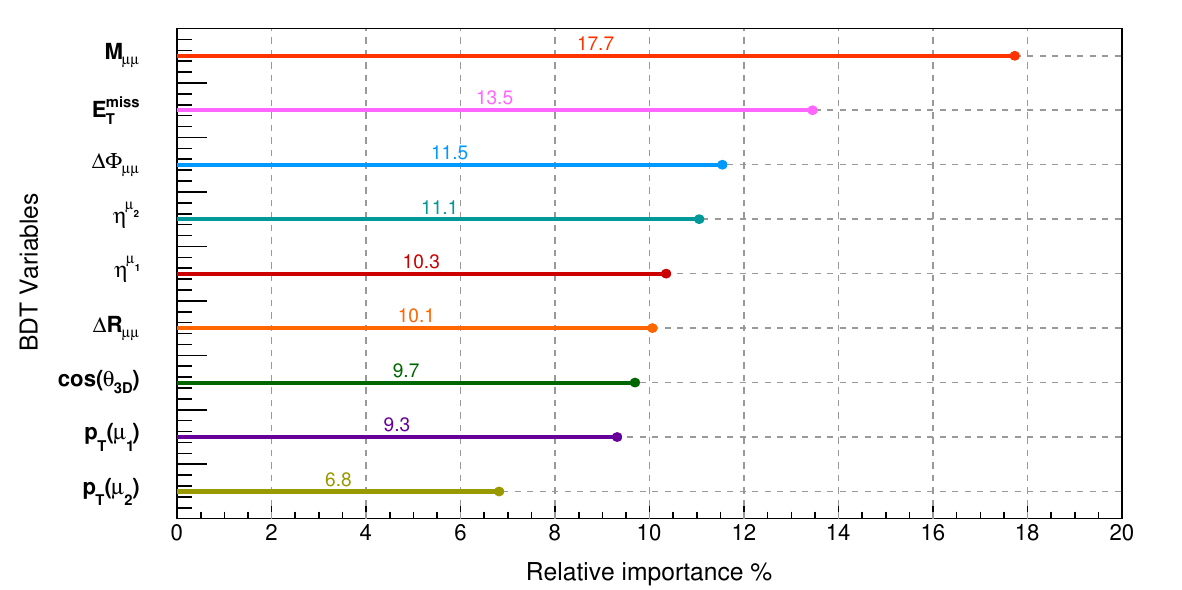}
     \caption{The relative importance of each of the BDT variables employed in the MVA for signal BMP6 versus all background at $\sqrt{s} = 500$ GeV.}
     \label{relimpo}
 \end{figure}

The relative importance of the variables used by the BDT (taking correlation into account) for BMP6 is illustrated in Figure \ref{relimpo}. The invariant mass proved to be an effective discriminant variable for this analysis (for $A^\prime$ masses that are not near the Z boson's mass).

\begin{figure}[h]
        \centering
        \includegraphics[width=1\linewidth]{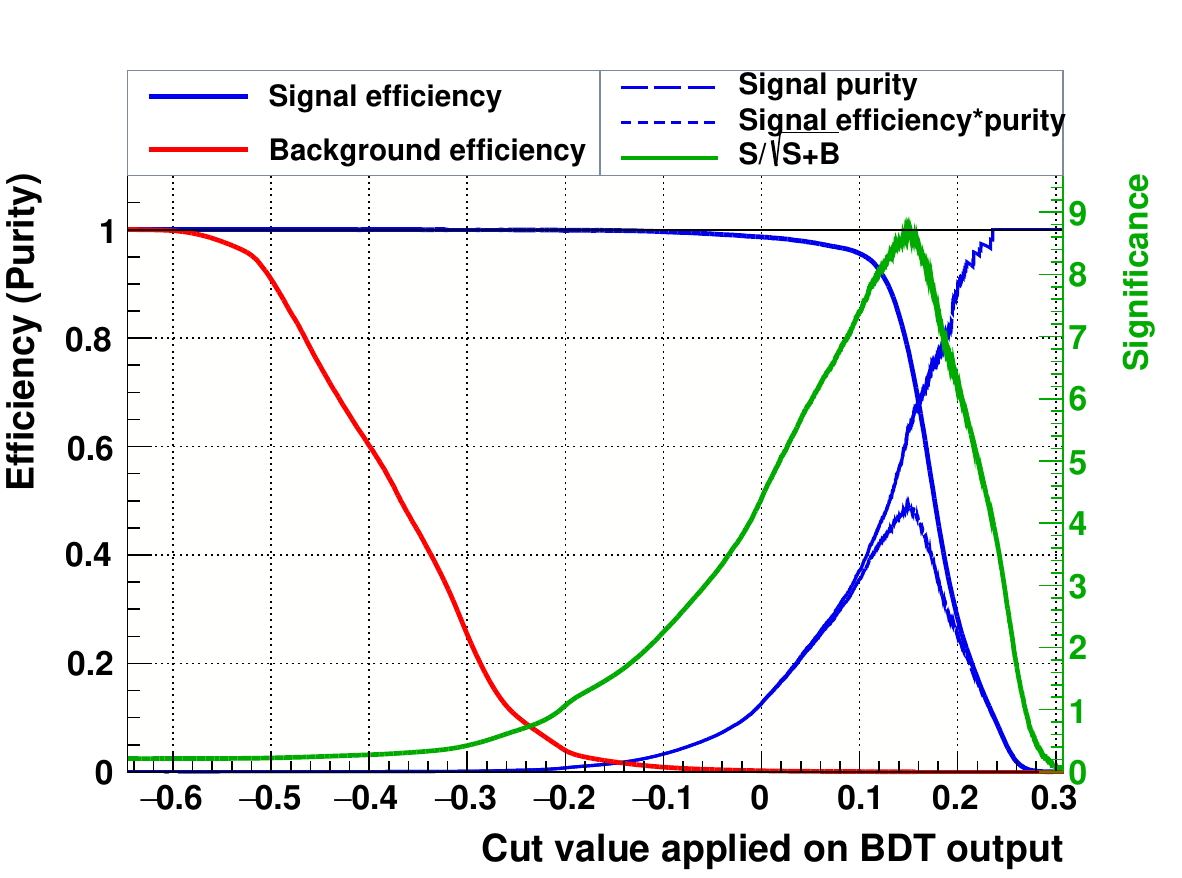}
        \caption{The BDT cut efficiency for BMP6 versus total SM background at $\sqrt{s} = 500$ GeV, $\mathcal{L}=500$ fb$^{-1}$. The optimal cut, at 0.1490, achieved a significance of 8.72}
  \label{BDTopt}
    \end{figure}

Using the number of the accepted events in the pre-selection denoted in Table \ref{preselec}, we obtained the significance after applying the BDT optimized cut for each BMP of the ECT signal at $\mathcal{L}=500$ fb$^{-1}$, and also measure the luminosity required to obtain the 5-sigma discovery ($\mathcal{L}_{\text{5$\sigma$}}^{\text{req}}$). The significance is calculated through \cite{TMVA}:

\begin{equation}
    S=\frac{\mathcal{N}_{S}}{\sqrt{\mathcal{N}_{S}+\mathcal{N}_{B}}}
    \label{eqn6}
\end{equation}

where $S$ is the significance, $\mathcal{N}_{S}$ and $\mathcal{N}_{B}$ represent, respectively, the number of signal and background events passing the BDT optimized cut.
Figure \ref{BDTopt} shows the BDT output for the reference point BMP6 where a significance of about 8.72 can be reached after applying the BDT optimized cut at 0.1490 at a luminosity of 500 fb${^{-1}}$.

\begin{table*}[]
    \centering
    \begin{adjustbox}{width=0.6\textwidth}
    \begin{tabular}{|c|c|c|c|c|c|c|c|}
        \hline
        BMPs & $\text{BDT}_{\text{opt}}$ & $\mathcal{N}_{S}(\epsilon_{S})$ & $\mathcal{N}_{B}(\epsilon_{B} \times 10^4)$ & $S (\sigma)$ & $\mathcal{L}_{\text{5$\sigma$}}^{\text{req}}$ ($fb^{-1}$) \\ \hline
        \hline
        
       1 & 0.1417 & 474 (0.884) & 127 (2.43) & 19.33 & 33.45 \\ \hline
       2 & 0.1375 & 400 (0.857) & 149 (2.85) & 17.07 & 42.9 \\ \hline
       3 & 0.1533 & 290 (0.786) & 114 (2.18) & 14.42 & 60.11 \\ \hline
       4 & 0.1516 & 224 (0.734) & 176 (3.37) & 11.20 & 99.65 \\ \hline
       5 & 0.1581 & 133 (0.538) & 879 (16.8) & 4.2 & 708.62 \\ \hline 
        \hline
        
       6 & 0.1490 & 120 (0.779) & 70 (1.34) & 8.72 & 163.27 \\ \hline
       7 & 0.1575 & 93 (0.727) & 82 (1.57) & 7.01 & 254.37 \\ \hline
       8 & 0.1646 & 73 (0.689) & 82 (1.57) & 5.87 & 362.77 \\ \hline
       9 & 0.1685 & 50 (0.575) & 79 (1.51) & 4.37 & 654.56 \\ \hline
       10 & 0.1608 & 35 (0.500) & 770 (14.8) & 1.22 & 8398 \\ \hline
        \end{tabular}
        \end{adjustbox}
    \caption{The calculated significance ($S(\sigma)$) along with the luminosity required to achieve a 5-$\sigma$ discovery ($\mathcal{L}_{\text{5$\sigma$}}^{\text{req}}$), for all signal benchmark points (BMPs) with $M_{\chi}$ set at 90 GeV, with coupling constants $\texttt{g}_{\eta} = 0.125$ and $\texttt{g}_{D} = 1.2$ at $\sqrt{s} =$ 500 GeV, and $\ \mathcal{L}=500$ fb$^{-1}$.}
      \label{results}
\end{table*}
\begin{figure}[!ht]
	\centering
    \begin{subfigure}{0.5\textwidth}
		\includegraphics[width=\textwidth]{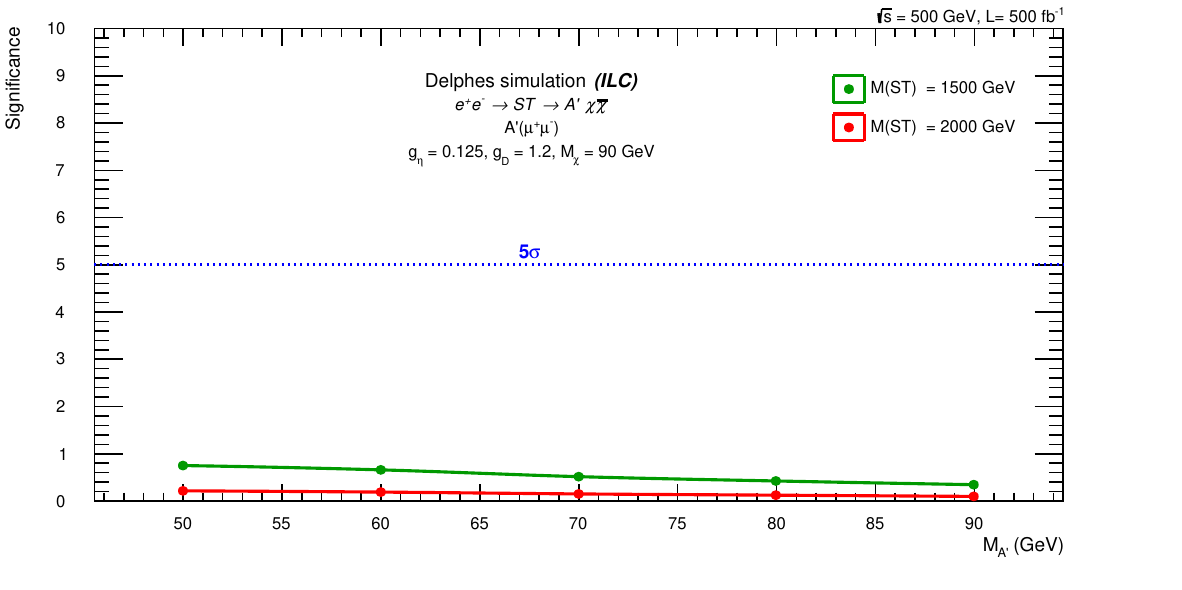}
		\caption{Before applying the BDT optimal cut.}
        \label{sig1}  
	\end{subfigure}
	\begin{subfigure}{0.5\textwidth}
		\includegraphics[width=\textwidth]{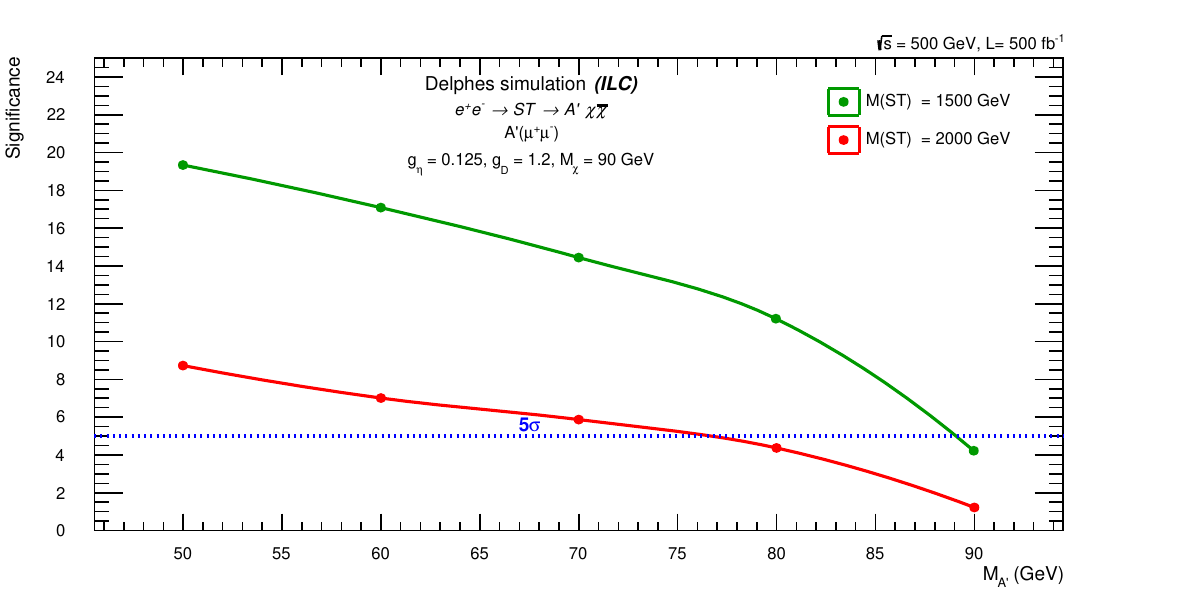}
		\caption{After applying the BDT optimal cut.}
        \label{sig2}
	\end{subfigure}
        \caption{The discovery potential for all signal benchmark points used in this study before (in \ref{sig1}) and after (in \ref{sig2}) applying the BDT optimal cut. The dotted horizontal blue line refers to the statistical 5-$\sigma$ value.}
  \label{Sig}
    \end{figure}

Table \ref{results} presents the full results of this analysis for all BMPs. Here $\text{BDT}_\text{opt}$ represents the BDT optimized cut for each signal BMP, $\mathcal{N}_S$ and $\mathcal{N}_{B}$ are the number of signal and SM background events surviving BDT optimized cut, while $S(\sigma)$ represents the statistical significance reached at $\mathcal{L}=$ 500 fb$^{-1}$, and $\mathcal{L}_{\text{5$\sigma$}}^{\text{req}}$ represents the required luminosity to achieve a 5-$\sigma$ discovery for each benchmark point, Finally, $\epsilon_{S}$ and $\epsilon_{B}$ indicate the acceptance efficiencies for the signal and the SM background events by the cut value $\text{BDT}_\text{opt}$.

From the values of $\mathcal{L}_{\text{5$\sigma$}}^{\text{req}}$, we can see that a 5-$\sigma$ discovery is easily achievable at the ILC at $\mathcal{L}=500$ fb$^{-1}$ with the use of the \texttt{TMVA} for most BMPs in the considered ECT signal at $\sqrt{s}$ = 500 GeV, indicating a strong discovery potential for this mass configuration. 
In contrast, a 5-$\sigma$ discovery for BMPs 5 and 9 would require a higher $\mathcal{L}$ but may still be easily achievable at a later stage at the ILC ($\sqrt{s}=1000$ GeV, $\mathcal{L}=1000$ fb$^{-1}$). This is not the case for BMP10. 
It would be extremely challenging to probe it at the ILC at any stage due to its extremely high $\mathcal{L}_{\text{5$\sigma$}}^{\text{req}}$, using the same features within the BDT. Higher mass points of ST were not added as they also have $\mathcal{L}_{\text{5$\sigma$}}^{\text{req}} >$ 1000 fb$^{-1}$ due to their lower cross-section values.
In addition to the low cross-section of BMP10, the BMPs 5 and 10 needed relatively higher $\mathcal{L}_{\text{5$\sigma$}}^{\text{req}}$ than others due to their invariant mass of 90 GeV being very close the mass of the Z boson which dominates in the invariant mass distribution of the background making them harder to discriminate, as also indicated from their relatively larger $\mathcal{N}_B$.

The findings of this analysis are highlighted by Figure \ref{Sig}, which demonstrates the discovery potential for all signal BMPs utilized in this study at the ILC. The statistical significance, calculated for all the analyzed signal points, is plotted against $M_{A^{\prime}}\ (GeV)$ at $M_{ST}=1500, 2000\ (GeV)$ representing all BMPs. This figure illustrates the pivotal role of the multivariate analysis, particularly through the BDT method, as seen in \ref{sig1} before applying the optimal BDT cut where there was no sensitivity at all, and in \ref{sig2} afterwards where most BMPs successfully reached a 5-$\sigma$ discovery, showcasing the ILC's impressive sensitivity to the ECT signal.
\section{Summary}
\label{section:conclusion}

Searching for the dark gauge boson A$^{\prime}$ at the International Linear Collider (ILC) was performed via its muonic decay within the framework of the ECT model.

Monte Carlo samples were used in this search, generated through \texttt{WHIZARD} package via the simulation of $e^+e^-$ interactions at $\sqrt{s} = $ 500 GeV, and $\mathcal{L}=$ 500 fb$^{-1}$, with polarization degrees of $P_{e^+} = 0.3$ and $P_{e^-} = -0.8$ for the positron and the electron beams, respectively.

Multiple benchmark points were used to test the signal of the bremsstrahlung scenario, using several A$^{\prime}$ masses, with fixed values of the coupling constants as $\texttt{g}_{D} = 1.2$ and $\texttt{g}_{\eta} = 0.125$ (for leptons) and fixed dark matter fermion mass ($M_{\chi}$ = 90 GeV), achieving the highest viable cross-section.

The multivariate analysis (MVA) enabled us to separate the signal from the SM background, and allowed us to achieve a 5-$\sigma$ discovery for most of the signal benchmark points at luminosities lower than the ILC's 500 fb$^{-1}$.

Within the context of the ECT model, achieving a 5-$\sigma$ discovery becomes quite challenging for mass values of $\text{M}_{ST} > 2000$ GeV and $\text{M}_{A^{\prime}} > 90$ GeV, especially with coupling constants set at $\texttt{g}_{D} = 1.2$ and $\texttt{g}_{\eta} = 0.125$ due to extremely low cross-section values. This would necessitate implementing additional new variables to enhance the separation between the signal and the SM background.


\begin{acknowledgments}
We would like to acknowledge the author of \cite{Einstein-Cartan}, Cao H. Nam, for providing the UFO files of the ECT model that was implemented in \texttt{WHIZARD} to generate the signal events.
\end{acknowledgments}

\textbf{Data Availability Statement:} This manuscript has no associated data. 
\nocite{*}


\end{document}